\documentclass{article}
\usepackage{amsmath,amsfonts,amssymb,amsthm,xspace,pstricks,longtable}
\newlength{\dinwidth}
\newlength{\dinmargin}
\makeatletter
\newif\if@fewtab\@fewtabtrue
\def\draftdate{\number\day.\number\month.\number\year\ \ \ \hourmin }
{\count255=\time\divide\count255 by 60
\xdef\hourmin{\number\count255}
\multiply\count255 by-60\advance\count255 by\time
\xdef\hourmin{\hourmin:\ifnum\count255<10 0\fi\the\count255}}
\def\ps@draft{\let\@mkboth\@gobbletwo
    \def\@oddhead{}
    \def\@oddfoot
       {\hbox to 7 cm{$\scriptstyle\bf Draft\ version:\ \draftdate$
       \hfil}\hskip -7cm\hfil\rm\thepage \hfil}
    \def\@evenhead{}\let\@evenfoot\@oddfoot}
\makeatother
\newcommand{\Field}[1]{\ensuremath{\mathbb{#1}}\xspace}

\newcommand{\Rn}{\Field{R}}

\newcommand{\Zn}{\Field{Z}}

\newcommand{\lb}[1]{\label{#1}}
\newcommand{\Eq}[1]{(\ref{#1})}
\newcommand{\ct}[1]{\cite{#1}}
\renewcommand{\[}{\begin{eqnarray}}
\renewcommand{\]}{\end{eqnarray}}
\newcommand{\nn}{\nonumber}
\newcommand{\non}{\nonumber \\ }

\newcommand{\een}{\end{enumerate}}
\newcommand{\ben}{\begin{enumerate}}
\newcommand{\gb}{\beta}

\newcommand{\gd}{\delta}
\newcommand{\gD}{\ensuremath{\Delta}\xspace}
\newcommand{\gep}{\epsilon}

\newcommand{\gL}{\ensuremath{\Lambda}\xspace}
\newcommand{\gm}{\mu}

\newcommand{\gr}{\rho}

\newcommand{\gt}{\theta}

\newcommand{\cC}{\mathcal{C}}

\newcommand{\cl}{\ensuremath{\ell}\xspace}
\newcommand{\cM}{\mathcal{M}}

\newcommand{\cW}{\mathcal{W}}

\newcommand{\fg}{\ensuremath{\mathfrak{g}}\xspace}

\newcommand{\fw}{\mathfrak{w}}
\newcommand{\fW}{\mathfrak{W}}

\renewcommand{\vec}[1]{{\boldsymbol{#1}}}
\newcommand{\vO}{\vec{0}}

\newcommand{\va}{\vec{a}}
\newcommand{\vb}{\vec{b}}

\newcommand{\vr}{\vec{r}}
\newcommand{\vrm}{{\vec{r}}_{-1}}
\newcommand{\vs}{\vec{s}}

\newcommand{\vv}{\vec{v}}

\newcommand{\vgb}{\vec{\gb}}
\newcommand{\vgd}{\vec{\gd}}

\newcommand{\vgL}{\vec{\gL}}

\newcommand{\vgr}{\vec{\gr}}
\newcommand{\vgt}{\vec{\gt}}

\newcommand{\mult}{\mathrm{mult}}

\DeclareMathOperator{\hgt}{ht}
\newcommand{\re}{\mathrm{e}}

\newcommand{\frc}[2]{\tfrac{#1}{#2}}
\newcommand{\8}{\ensuremath{E_8}\xspace}
\newcommand{\9}{\ensuremath{E_9}\xspace}
\newcommand{\0}{\ensuremath{E_{10}}\xspace}

\newcommand{\X}{\!\cdot\!}

\newcommand{\latt}[1]{\ensuremath{I\hspace{-.2em}I_{#1,1}}\xspace}

\newcommand{\Lie}[1]{\ensuremath{\fg_{\latt{#1}}}\xspace}

\begin{document}
\def\draft{\pagestyle{draft}\thispagestyle{draft}
\global\def\draftcontrol{1}}
\global\def\draftcontrol{0}
%------------------- switch on/off draft version -----------------------
%\draft
%-----------------------------------------------------------------------
\arraycolsep3pt

\thispagestyle{empty}
\begin{flushright} hep-th/9705144
               \\  IASSNS-HEP-97-53
               \\  AEI-037
\end{flushright}
\vspace*{2cm}
\begin{center}
 {\LARGE \sc On the Imaginary Simple Roots of\\[1ex]
             the Borcherds Algebra $\Lie9$%
%            \footnote{submitted to \emph{???}}
  }\\
 \vspace*{1cm}
 {\sl
     Oliver B\"arwald\footnotemark[1]$^,$\footnote[2]{Supported by
     \emph{Gottlieb Daimler- und Karl
                   Benz-Stiftung} under Contract No.\ 02-22/96},
     Reinhold W.~Gebert\footnotemark[3]$^,$%
      \footnote[4]{Supported by \emph{Deutsche Forschungsgemeinschaft}
                   under Contract No.\ DFG Ge 963/1-1} and
     Hermann Nicolai\footnotemark[5]$^,$%
      \footnote[6]{Supported in part by EU Contract FMRX-CT96-0012} \\
 \vspace*{6mm}
     \footnotemark[1]
     Department of Mathematics, King's College London\\
     Strand, London WC2R 2LS, Great Britain\\
 \vspace*{3mm}
     \footnotemark[3]
     Institute for Advanced Study, School of Natural Sciences\\
     Olden Lane, Princeton, NJ 08540, U.S.A.\\
 \vspace*{3mm}
     \footnotemark[5]
     Max-Planck-Institut f\"ur Gravitationsphysik\\
     Albert-Einstein-Institut \\
     Schlaatzweg 1, D-14473 Potsdam, Germany} \\
 \vspace*{6mm}
\ifnum\draftcontrol=1{\Large\bf Draft version: \draftdate \\}
                     \else{8 September 1997 \\}\fi
 \vspace*{1cm}
\begin{minipage}{11cm}\footnotesize
\textbf{Abstract:}
In a recent paper \ct{BaGeGuNi97} it was conjectured that the
imaginary simple roots of the Borcherds algebra $\Lie9$ at level 1 are
its only ones. We here propose an independent test of this conjecture,
establishing its validity for all roots of norm $\geq -8$.
However, the conjecture fails for roots of norm $-10$ and beyond,
as we show by computing the simple multiplicities down to norm $-24$,
which turn out to be remakably small in comparison with the corresponding
\0 multiplicities. Our derivation is based on a modified denominator formula
combining the denominator formulas for \0 and $\Lie9$, and provides
an efficient method for determining the imaginary simple roots.
In addition, we compute the $\0$ multiplicities of all roots up
to height 231, including levels up to $\cl =6$ and norms $-42$.
\end{minipage}
\end{center}
\setcounter{footnote}{0}
%\newpage

%-----------------------------------------------------------------------
\section{Introduction} \lb{sec:Int}
%-----------------------------------------------------------------------
In this paper we begin a systematic study of the simple imaginary roots
of the Borcherds Lie algebra $\Lie9$ and propose a new method to compute
their (simple) multiplicities, enabling us to test the conjecture made
in \ct{BaGeGuNi97} concerning the set of imaginary simple roots of
$\Lie9$. Let us recall that $\Lie9$ is the Lie algebra of physical
states of a subcritical bosonic string fully compactified on the even
self-dual Lorentzian lattice $\latt9$, which coincides with the root
lattice $Q(\0)$ of the hyperbolic Kac--Moody algebra $\0$; this
lattice is spanned by the simple roots $\vrm,\vr_0,\vr_1,\ldots,\vr_8$
of \0 (alias the real simple roots of $\Lie9$). Our primary
motivation for investigating the root system of $\Lie9$ is to better
understand the hyperbolic algebra $\0$, which is the maximal
Kac--Moody algebra contained in $\Lie9$:
\[
\0\subset\Lie9. \lb{E10} \]
As explained in \ct{BaGeGuNi97}, the difficulties of understanding
hyperbolic Kac--Moody algebras on the one hand and Borcherds algebras
(also called generalized Kac--Moody algebras \ct{Borc88}) on the other
hand are to some extent complementary: while a Lorentzian Kac--Moody
algebra has a well-understood root system, but the structure of the
algebra and its root spaces (including their dimensions!) is very
complicated, Borcherds algebras may possess a simple realization in
terms of physical string states, but usually have a very complicated
root system due to the appearance of imaginary (i.e.\ non-positive
norm) simple roots. The Chevalley generators corresponding to
imaginary simple roots of $\Lie9$ are needed to complete the
subalgebra \0 to the full Lie algebra of physical states. This can be
seen by decomposing the vector space
\[
\cM := \Lie9 \ominus \0     \lb{cM}
\]
into an infinite direct sum of ``missing modules'' all of which are
highest or lowest weight modules w.r.t.\ the subalgebra $\0$ (see
\ct{BaGeGuNi97,Juri97}). This implies that all of $\Lie9$ can be
generated from the highest and lowest weight states by the action of
(i.e. multiple commutation with) the \0 raising or lowering
operators. However, not all the lowest weight states in $\cM$
correspond to imaginary simple roots of $\Lie9$. This is because the
commutation of two lowest weight states yields another lowest weight
state; yet it is only those lowest weight states which cannot be obtained
as multiple commutators of previous states and which must therefore be
added ``by hand'' that will give rise to new imaginary simple roots.
Complete knowledge of the imaginary simple roots of $\Lie9$
is thus tantamount to understanding the hyperbolic Kac--Moody
algebra $\0$ (or at least its root multiplicities).

Let us pause for a moment to rephrase these statements in string
theory language. As has been shown in \ct{GeNiWe96}, commutation of
two physical string states in these completely compactified string
models is equivalent to tree-level scattering. So, starting with a set
of ten fundamental tachyons (associated with the real simple roots),
we generate by multiple scattering an infinite set of physical string
states at arbitrary mass level. By construction, this set is just the
hyperbolic algebra \0, and it is easy to see that it contains all
tachyonic and all massless states as these can be produced by
elementary scattering processes. By contrast, \0 does not exhaust the
massive states because not all of these can be obtained by scattering
\0 states of lower mass. To be sure, \0 does act via the adjoint
action on all physical states, i.e., we can scatter two states only
one of which belongs to \0 to get another state, which is also not in
$\0$. Therefore, the remaining (``missing'', or ``decoupled'')
part of the spectrum can be decomposed into \0 representations.
In order to identify the
pertinent highest or lowest weight states, the strategy is to pick
suitable missing string states of lowest mass and add them as
extra Lie algebra elements to \0. Since the momenta of these states
have negative norm\footnote{By the ``norm'' of a root $\vgL$ we mean
the (Minkowskian) scalar product $\vgL^2$.} this corresponds to
adjoining timelike simple roots to \0. These simple roots generically
come with multiplicities bigger than one because massive string states
have additional polarization degrees of freedom, whereas the tachyons
are scalars, and the real simple roots consequently have multiplicity
one always. Following \ct{BaGeGuNi97} we will designate the simple
multiplicity of an imaginary simple root $\vgL$ by $\gm(\vgL)$; this
simple multiplicity $\gm(\vgL)$ must not be confused with the
multiplicity $\mult (\vgL)$ of $\vgL$ as a root of $\0$ or with the
multiplicity $\dim\Lie9^{(\vgL)}$ of $\vgL$ as a root of $\Lie9$.
Therefore, the Lie algebra of all physical states is no longer a
Kac--Moody algebra since the Cartan matrix may now have negative
integers on the diagonal. In general, the above procedure has to be
repeated an infinite number of times because by scattering the
adjoined massive states with \0 states, we still do not exhaust the
whole spectrum.

So far, there are only a few string models for which the root system
of the associated Borcherds algebra has been completely analyzed, and
for which a complete set of imaginary simple roots associated with
extra string states, including their multiplicities, has been
identified. Celebrated examples are the fake monster \ct{Borc90} and
the monster Lie algebra \ct{Borc92}, which are related to a toroidal
and an orbifold compactification of the 26-dimensional bosonic
string. In \ct{BaGeGuNi97}, an infinite set of level-1 imaginary
simple roots (with exponentially growing, known multiplicities) for
the Borcherds algebra $\Lie9$ was found and it was conjectured that
this set should exhaust all of them. The results of this paper
disprove the conjecture and show that the structure of the $\Lie9$
root system is more involved than originally thought. In establishing
these results, we are led to explore the multiplicities and simple
multiplicities of these algebras much further than has been done
before.  Our calculations of the simple root multiplicites are based
on a new denominator formula which combines the known denominator
formulas for \0 and $\Lie9$. Although not as efficient as the Peterson
recursion formula (which appears to have no analog for simple
multiplicities), this formula does simplify the computations
substantially and allows us to evaluate the simple multiplicities down
to norms $-24$. Since we have made no attempt to optimize our computer
program (using the symbolic algebra
system \texttt{Maple V}) with regard to speed, it is
quite conceivable that the calculations can be carried even
further. As an important by-product of this investigation we have
determined the \0 root multiplicities up to height 231 including
levels up to $\cl =6$ and norms down to $-42$, because these numbers
are needed as an input in our modified denominator formula. Since
these results may also be of use in other contexts, we have tabulated
them separately in appendix A.

Although the ultimate pattern underlying the multiplicities
remains elusive, and despite the failure of our original conjecture,
our results do exhibit some intriguing features. In particular, we would
like to draw attention to the fact that, as far as we have
computed them, the simple multiplicities come out to be remarkably
small, both in comparison with the $\0$ multiplicities and with
the number of decoupled states. For instance, at level $\cl=4$,
we find
$$
\mu (\vgL_3)=2,
$$
whereas the \0 multiplicity of $\vgL_3$ is given by
$$
\mult (\vgL_3) = 1044218,
$$
and the number of associated decoupled states is equal to
$$
\gD(\vgL_3)= 278125.
$$
This behavior is to be contrasted with the gnome Lie algebra \Lie1 for
which the simple root multiplicities and the root space dimensions are
of the same order of magnitude \cite{BaGeGuNi97}. On the other hand,
the pattern is clearly more irregular than that of the fake monster
algebra \Lie{25} whose simple multiplicities either vanish, or are
uniformly equal to $24$. In fact, at this point, we cannot even
exclude the possibility of a ``chaotic'' behavior at yet larger
(negative) norms and higher levels --- after all, \0 is defined by
means of a recursive procedure just like simple chaotic systems.
Being optimistic and barring such pathologies, the supreme challenge
is now to unveil the secret behind the numbers we have found; and,
whatever hypothesis is profferred to explain $\0$, it must be tested
against these numbers.

%-----------------------------------------------------------------------
\section{A modified denominator formula} \lb{sec:Mod}
%-----------------------------------------------------------------------
We first summarize our notations and conventions for $\0$, mostly following
\ct{KaMoWa88} to which we refer the reader for further details.
The real simple roots $\vr_i$ and the fundamental weights $\vgL_i$
are labeled in accordance with the Coxeter--Dynkin diagram:
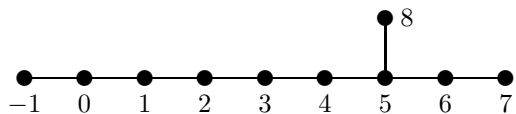
\[ \unitlength1mm
   \begin{picture}(66,15)
   \multiput(1,5)(8,0){9}{\circle*{2}}
   \put(49,13){\circle*{2}}
   \multiput(2,5)(8,0){8}{\line(1,0){6}}
   \put(49,6){\line(0,1){6}}
   \put(1,.5){\makebox[0pt]{$-1$}}
   \put(9,.5){\makebox[0pt]{$0$}}
   \put(17,.5){\makebox[0pt]{$1$}}
   \put(25,.5){\makebox[0pt]{$2$}}
   \put(33,.5){\makebox[0pt]{$3$}}
   \put(41,.5){\makebox[0pt]{$4$}}
   \put(49,.5){\makebox[0pt]{$5$}}
   \put(57,.5){\makebox[0pt]{$6$}}
   \put(65,.5){\makebox[0pt]{$7$}}
   \put(51,12){\makebox[0pt][l]{$8$}} \end{picture}  \nn  \]
from which the \0 Cartan matrix $A_{ij}$ can be easily read off.
The root $\vrm$ which extends the affine subalgebra \9 to the full
hyperbolic algebra \0 will be referred to as the over-extended root.
The level \cl of a root $\vr$ in the fundamental Weyl chamber is defined by
\[
\cl:=-\vgd\X\vr   \lb{levl}
\]
where $\vgd$ denotes the affine null root. The fundamental Weyl
chamber $\cC$ is the positive convex cone in $\latt9$ generated by
the fundamental weights $\vgL_i= - \sum_j (A^{-1})_{ij}\vr_j >0$,
obeying $\vgL_i\X\vr_j=-\gd_{ij}$ (this is the only place where we
deviate from the conventions of \ct{KaMoWa88}). So we have, for
instance, $\vgL_{-1}=\vgd$ and $\vgL_0=\vrm + 2\vgd$, etc.; since
$\vgL_{-1}^2 =0$ and $\vgL_i^2<0$ for $i\geq 0$, $\cC$ lies inside the
forward lightcone and touches it at one edge. Acting on $\cC$ with all
elements of the \0 Weyl group and taking the closure of the resulting
set, one obtains the so-called Tits cone which coincides with the full
forward light-cone containing all imaginary roots \ct{Kac90}. The Weyl
chamber $\cC$ contains in particular the imaginary simple roots which
must be adjoined to complete \0 to the full algebra $\Lie9$ of
physical states. For the determination of root multiplicities it is
therefore sufficient to restrict attention to roots in the fundamental
Weyl chamber $\cC$; for a given root norm the computation can thus be
reduced to a a finite number of checks.

In \ct{BaGeGuNi97} a complete characterization of all level-1 imaginary
simple roots of $\Lie9$ and their multiplicities was given: the
missing lowest weight states are just the purely longitudinal physical
states with momenta $\vrm + N\vgd$ for $N\geq 2$. The multiplicities
of these simple roots are given by $\gm (\vrm + N\vgd) = \pi_1 (N)$,
where
\[
\pi_d(n):= p_d(n)-p_d(n-1)  \lb{Part1}
\]
with
\[
\sum_{n=0}^\infty p_d(n)q^n = \prod_{n\geq 1}(1-q^n)^{-d}. \lb{Part2}
\]
One would expect the structure to be far more involved at higher
levels, but the explicit calculations in \ct{BaGeGuNi97} revealed that
there were no imaginary simple roots $\vs$ at level 2 with $\vs^2\geq
-6$.  This unexpected result prompted the conjecture that the level-1
roots of $\Lie9$ are in fact the only imaginary simple roots of
$\Lie9$, or, equivalently, that
the set of missing lowest weight states is the free Lie algebra
generated by the purely longitudinal states at level 1. The evidence
presented in \ct{BaGeGuNi97} was based on computer calculations of
commutators of certain level-1 states but this method becomes
impractical beyond the examples studied there because of the rapidly
increasing algebraic complexity as the root norms become more
negative. We here present an independent test via a
modified denominator formula, enabling us to carry the checks much
further, even without use of a computer. This new formula combines the
\0 denominator formula with the one for $\Lie9$, and is therefore
sensitive only to the ``difference'' of these two algebras.

The denominator formula for \0 reads (see e.g. \ct{Kac90})
\[
\prod_{\vr\in\gD_+}(1-\re^\vr)^{\mult(\vr)}
=\sum_{\fw\in\fW}(-1)^\fw\re^{\fw(\vgr)-\vgr},  \lb{Nenner1}
\]
where $\gD_+$ are the positive roots of \0 and $\vgr$ is the \0 Weyl
vector, i.e., $\vgr\X\vr_i= -\tfrac12\vr_i^2$ for $i=-1,0,1,\ldots,8$.
The determination of the root multiplicities
\[
\mult (\vr):= \dim \, \0^{(\vr)} \lb{mult}
\]
at arbitrary level $\cl$ remains an unsolved problem for \0 (and, more
generally, for any hyperbolic Kac--Moody algebra): closed formulas
exist only for levels $|\cl | \leq 2$ \ct{KaMoWa88}, and, albeit in
implicit form, for $\cl =\pm 3$ \ct{BauBer97}. The denominator
formula relates the infinite product over all positive roots to an
infinite sum over the Weyl group $\fW\equiv \fW(\0)$ generated by the
reflections w.r.t.\ the real simple roots. In principle, all root
multiplicities can be determined from it by multiplying out the
l.h.s.\ and comparing the resulting expressions term by term, but in
practice this method reaches its limits rather quickly. However, one
can derive from \Eq{Nenner1} the so-called Peterson recursion
formula (see e.g.\ \ct{KMPS90}), which can be implemented on a computer.
Because, to the best of our knowledge, explicit tables of \0 multiplicities
available in the literature stop at $|\cl |=2$ \ct{KaMoWa88} and
the actual numbers are needed in our calculation, we have computed
the \0 multiplicities of all roots up to height 231 and levels $\leq 6$
by putting the Peterson formula on a computer. Readers may notice
that, beyond $\cl =2$, these multiplicities are no longer functions
of the norm alone, as was still the case for the level 2 roots
(so Murphy's law has struck again).

For the Borcherds algebra $\Lie9$, the denominator formula must be
amended in two ways: firstly, the \0 multiplicities $\mult(\vr)$
are replaced by the corresponding numbers of physical states
\[
\dim \, \Lie9^{(\vr)}= \pi_9(1-\frc12 \vr^2) \geq \mult(\vr), \lb{dim}
\]
and secondly, the r.h.s.\ must be supplemented by extra terms due
to the imaginary simple roots. The modified denominator formula
reads \ct{Borc88}
\[
\prod_{\vr\in\gD_+}(1-\re^\vr)^{\pi_9(1-\frc12 \vr^2)}
 &=&\sum_{\fw\in\fW}(-1)^\fw\re^{\fw(\vgr)-\vgr}
 \sum_{\vs}\gep(\vs)\re^{\fw(\vs)},    \lb{denom}
\]
where $\gep(\vs)$ is $(-1)^n$ if $\vs$ is a sum of $n$ distinct
pairwise orthogonal imaginary simple roots of $\Lie9$, and zero otherwise.
As already pointed out the candidates for imaginary simple
roots are all lattice points in the fundamental Weyl chamber $\cC$.
Because all massless physical string states belong to \0, there
are no lightlike simple roots. Consequently, the imaginary simple
roots are all timelike, and we therefore conclude that there
are no pairwise orthogonal imaginary simple roots. Hence
\[
\sum_{\vs}\gep(\vs)\re^{\fw(\vs)}
&=& 1-\sum_{\substack{\vgL\in\cC\cap\latt9 \\ \vgL^2\le-2}}
      \gm(\vgL)\re^{\fw(\vgL)} \non
&=& 1-\sum_{N=2}^\infty \pi_1(N) \re^{\fw(\vrm + N\vgd)}
       + \dots\,, \lb{Nenner2}
\]
where, in the second line of this equation, we have made use of our
knowledge of the simple roots at level 1 and their multiplicities; the dots
stand for possible contributions from higher-level imaginary simple roots.
Unfortunately, there seems to be no analog of the Peterson recursion
formula that would allow a (comparatively) quick determination of the
imaginary roots and their simple multiplicities. The proof
of this formula crucially relies upon the existence of a Weyl vector
with $\vgr\cdot\vs = - \frc12 \vs^2$ for {\em all} simple roots $\vs$:
while this requirement is met by the ten simple roots of
$\0$ it fails already for the level-1 imaginary simple roots
$\vs=\vrm +N\vgd$, as one can easily check. Consequently, no Weyl
vector exists for \Lie9 and we have to seek another way of simplifying
the denominator formula in order to test our conjecture.

The idea is to modify the formula in such a way that it ``measures''
only the corrections that arise when enlarging \0 to the full Lie
algebra $\Lie9$ of physical states. To this aim let us introduce the
difference between the \Lie9 and the \0 multiplicities, i.e., the
number of decoupled (missing) states associated with the root $\vr$,
\[
\gD (\vr):= \pi_9 (1 -\frc12 \vr^2) - \mult (\vr) \qquad(\geq 0). \lb{D}
\]
So, using the known results in for $|\cl|\leq 2$, we have
\[
\gD (\vr)=
\begin{cases}
0 & \text{for $\cl=0$}, \\
\pi_9 (1 -\frc12 \vr^2) - p_8 (1 -\frc12 \vr^2 ) & \text{for $\cl=1$}, \\
\pi_9 (1 -\frc12 \vr^2) - \xi (3 -\frc12 \vr^2 ) & \text{for $\cl=2$},
\end{cases}
\]
where the function $\xi(n)$ was defined and tabulated in
\ct{KaMoWa88} (notice the accidental equality $p_8(3)= \xi(5)$).
The explicit \0 multiplicities beyond level 2 which we need
have been collected in the table of Appendix A, where we also
list the relevant values for $\gD(\vr)$.

Inserting the \0 denominator formula into the one for $\Lie9$,
we obtain the following formula after a little algebra:
\[  \bigg(\sum_{\fw\in\fW} (-1)^\fw \re^{\fw(\vgr)}\bigg)
    \bigg(\prod_{\vr\in\gD_+} (1-\re^\vr)^{\gD(\vr)} -1\bigg)
&=& -\sum_{\fw\in\fW}
     \sum_{\substack{\vgL\in\cC\cap\latt9 \\ \vgL^2\le-2}}
      (-1)^\fw \gm(\vgL)\re^{\fw(\vgr + \vgL)} \non
&=& - \sum_{\fw\in\fW} \sum_{N=0}^\infty (-1)^\fw  \, \pi_1 (N+2)
     \re^{\fw( \vgr + \vgL_0 + N\vgL_{-1})} + \dots\, . \lb{Nenner3}
\]
The dots stand for the level $\cl\geq 2$ contributions which
were conjectured to vanish in \ct{BaGeGuNi97}. We will show explicitly
how the conjecture fails by exhibiting non-zero contributions of this type.

One advantage of this formula is the absence of terms without
$\vr_{-1}$ on the l.h.s.\ since for all such roots we have
$\gD(\vr)=0$; thus, the \9 part of the denominator formula has already
been factored out in \Eq{Nenner3}. Given the \0 root multiplicities it
allows us to determine the simple roots together with their simple
multiplicities rather efficiently, as we will demonstrate in the
following section. The analysis of \Eq{Nenner3} can be considerably
simplified by restricting oneself on the r.h.s.\ to roots in the
fundamental Weyl chamber $\cC$. To see this, we note that in terms of
the fundamental weights we have $\vgr=\sum_{j=-1}^8 \vgL_j$, so that
$\vgr + \vgL\in \cC$ for all $\vgL\in\cC$, and that for any $\fw\neq
1$ the vector $\fw(\vgr + \vgL)$ lies outside the fundamental Weyl
chamber since no fundamental reflection leaves $\vgr$ invariant. With
this observation, the sum over the Weyl group on the r.h.s.\ can be
disregarded.

The general procedure for evaluating the new denominator formula
is then as follows. Let us fix a dominant integral level-\cl weight
$\vgL\in\cC$ for which $\vgL=\sum m_j \vgL_j$ with $m_j\geq 0$.
We wish to determine the coefficient of $\re^{\vgL}$ on the l.h.s.\
of formula \Eq{Nenner3}.  To do so, we must first look
for all possible decompositions $\vgr + \vgL = \fw(\vgr) + \vv$
with $\vv\in Q_+(\0)$. The reason why we cannot drop the
sum over the Weyl group on the l.h.s., even if we consider only
terms in the fundamental Weyl chamber on the r.h.s., is that in this
decomposition neither $\fw(\vgr)$ nor $\vv$ will in general be in
the fundamental Weyl chamber even if their sum is. Now, for
$$
\fw(\vgr)= \vgr + \va
$$
we have $\va>0$ unless $\fw =1$; this follows from the fact that the
Weyl vector is a dominant \0 weight. From the preservation of the scalar
product and basic properties of the Weyl vector we deduce that (for
$\va\neq\vO$)
$$
\va^2 = -2\vgr\X\va = 2\hgt(\va) > 0,
$$
where ``$\hgt$'' denotes the height of the root, and
$$
\vv^2 = \vgL^2 + 2 \big(\hgt(\va)-\va\X\vgL\big).
$$
Note that for $\vgL\in\cC$ we also have $\va\X\vgL\leq 0$ for any
positive $\va$. These simple relations severely constrain the possible
$\va$'s that must be taken into account: since repeated Weyl
reflections will increase the height, there are only finitely many
terms which can contribute for any given $\vgL$. Having found all
possible $\vv=\vgL-\va\in Q_+(\0)$ that can appear, the next problem
will be to calculate the coefficient of $\re^{\vv}$ arising from the
product over the positive roots. For this we must find all
decompositions $\vv = \sum_j \vv_j$ with $\vv_j\in\gD_+$. Some care
must be exercised with the various minus signs arising from the Weyl
reflections as well as from the binomial expansions of the factors in
the product over the positive roots. In particular, we have to know
all multiplicities of the relevant positive roots. At higher level,
this causes the extra complication of determining in which \0 Weyl
orbits these roots lie. Given any positive root $\vr$, this amounts to
reflecting it by use of the Weyl group into the fundamental
chamber. Although there seem to be no general results available we
have found the following method, due to J. Fuchs \ct{Fuchs}, to be
very efficient. One starts by rewriting the root in the basis of
fundamental weights, i.e., $\vr= \sum_{i}m_i\vgL_i$ for $m_i\in
\Zn$. Since $\vr\notin\cC$ by assumption, at least one of the
coefficients $m_i$ is negative. Choose a negative coefficient with the
largest absolute value $m_k$, say, and apply the $k$th fundamental
Weyl reflection to the root. We obtain $\fw_k(\vr)=
\sum_{i}\fw_k(m_i)\vgL_i$ with $\fw_k(m_i):= m_i-m_kA_{ki}$, so that
the coefficient of $\vgL_k$ is now $-m_k>0$.  The next step is to
determine again the most negative coefficient, to apply the
corresponding Weyl reflection and so on. This algorithm always
terminates after a finite number of steps.

%-----------------------------------------------------------------------
\section{Sample calculations for $\vgL^2 \geq -10$} \lb{sec:Det}
%-----------------------------------------------------------------------
Let us now illustrate how the calculation works in detail for
some simple examples for which the new denominator formula can
be evaluated by hand. This is certainly the case for roots $\vgL\in\cC$
with $\vgL^2\geq -10$, and perhaps beyond; however, the
combinatorial complexities, and thus the possible sources
of errors, increase rapidly for large negative $\vgL^2$, and
we have therefore preferred to let the computer do the rest of
the calculation, see the following section.

For the level-1 roots the required computations are quite
straightforward, as we need only consider $\fw\in\fW(\9)$ and make
use of the fact that the \9 Weyl orbit of $\vrm +N\vgd$ consists of
all elements $\vrm + (N+\frc12 \vb^2)\vgd + \vb$ with $\vb\in Q(\8)$.
It is then not difficult to check the validity of \Eq{Nenner3}
for roots $\vgL=\vgL_0 + (N-2) \vgL_{-1} = \vrm + N\vgd$
to large $N$. However, since we anyhow know the formula to
be correct at level 1, we refrain from giving further details.
As regards the level-2 roots of norm $\geq -6$, our
calculation will just confirm the conclusions reached in
\ct{BaGeGuNi97}, whereas for norms $-8$ and $-10$ our results are new
(and unlikely to be obtainable by the methods of \ct{BaGeGuNi97}).
We will show that all higher-level terms on the r.h.s.\ of Eq.\
\Eq{Nenner3} down to norm squared $-8$ are absent, in agreement with
the conjecture of \ct{BaGeGuNi97}; the relevant roots in
$\cC$ are $\vgL_7, \vgL_1, 2\vgL_0$ and $\vgL_7+\vgd$, all of level 2.
There are two norm $-10$ roots in $\cC$, namely $\vgL_1 +\vgd$, of
level 2, and $\vgL_8$, of level 3 (all other level $\geq 3$ roots in
$\cC$ have norm $\leq -12$); for these, we will find a non-vanishing
result, refuting the conjecture of \ct{BaGeGuNi97}. Further
counterexamples will be provided in the next section.

In the actual calculations we will need to determine to which
Weyl orbit a given root belongs, and whenever referring to a
root lying in a certain Weyl orbit we have checked this by Fuchs'
algorithm. For small norms this is not really necessary, if there is
only one Weyl orbit; for instance, there is only one orbit
$\fW(\vgL_0)$, for roots with $\vr^2=-2$ which has $\gD(\vgL_0)=1$.
For roots with $\vr^2=-4$ there are two Weyl orbits,
$\fW(\vgL_0+\vgd)$ and $\fW(\vgL_7)$, which happen to yield the same
numbers $\gD(\vgL_0+\vgd)= \gD(\vgL_7)= 9$. In fact, all Weyl orbits
of level-2 roots in $\cC$ with the same norm have the same value for
$\gD$ because the level-2 multiplicities (described by the function
$\xi$) depend only on the norm.  The combinatorial prefactors below
arise from the combinatorics of the indices $i,j,\ldots$ and are most
conveniently determined by inspection of the Coxeter--Dynkin diagram.
Relations such as $0\le\va^2<-\vgL^2+2\va\X\vgL$ show that we have listed
all nonzero contributions to the simple multiplicities of the roots under
consideration. Finally, an important consistency check on the
calculation is that the non-zero coefficients must come out to be
non-positive due to the absence of pairwise orthogonal simple
imaginary roots. This applies in particular to the large (negative)
norm roots to be analyzed in the next section, where the final result
is obtained as an alternating sum of huge contributions.

\ben
\item $\vgL^2=-4$, i.e.\ $\vgL=\vgL_7$.
 \ben
 \item $\vgL\X\va=0$:
  \ben
  \item $\va=\vO$
   $\quad\Longrightarrow\quad -\gD(\vgL_7)=-9$;
  \item $\va^2=2$, i.e.\ $\va=\vr_i$ for $i\neq7$
   $\quad\Longrightarrow\quad 9\X(-1)^2\gD(\vgL_0)=9$.
  \een
 \een
In total, this gives $-9+9=0$ for the simple multiplicity.
\item $\vgL^2=-6$, i.e.\ $\vgL=\vgL_1$.
 \ben
 \item $\vgL\X\va=0$:
  \ben
  \item $\va=\vO$
   $\quad\Longrightarrow\quad -\gD(\vgL_1)=-53$;
  \item $\va^2=2$, i.e.\ $\va=\vr_i$ for $i\neq1$
   $\quad\Longrightarrow\quad
   9\X(-1)^2\gD(\vgL_7)=81$;
  \item $\va^2=4$, i.e.\ $\va=\vr_i+\vr_j$ for $i,j\neq 1$ and
   $\vr_i\X\vr_j=0$
   $\quad\Longrightarrow\quad 29\X(-1)^3\gD(\vgL_0)=-29$;
  \een
 \item $\vgL\X\va=-1$:
  \ben
  \item $\va^2=2$, i.e.\ $\va=\vr_1$
   $\quad\Longrightarrow\quad (-1)^2\gD(\vgL_0)=1$.
  \een
 \een
In total, this gives $-53+81-29+1=0$ for the simple
multiplicity.\footnote{These two examples amply demonstrate the power
 of formula \Eq{Nenner3}: the necessary commutator calculations
 in \ct{BaGeGuNi97} needed to reach the same conclusion required two
 hours of CPU time!}
\item $\vgL^2=-8$, i.e.\ $\vgL=2\vgL_0$ or $\vgL=\vgL_7+\vgd$. \\[1.5ex]
Let $\vgL=2\vgL_0$.
 \ben
 \item $\vgL\X\va=0$:
  \ben
  \item $\va=\vO$
   $\quad\Longrightarrow\quad -\gD(2\vgL_0)=-246$;
  \item $\va^2=2$, i.e.\ $\va=\vr_i$ for $i\neq0$
   $\quad\Longrightarrow\quad
   (-1)^2[\gD(\vgL_0+2\vgd)+8\gD(\vgL_1)]=478$, \\
   since $\vgL-\vrm\in\fW(\vgL_0+2\vgd)$
   and $\vgL-\vr_1,\ldots,\vgL-\vr_8\in\fW(\vgL_1)$;
  \item $\va^2=4$, i.e.\ $\va=\vr_i+\vr_j$ for $i,j\neq 0$ and
   $\vr_i\X\vr_j=0$
   $\quad\Longrightarrow\quad
   29\X(-1)^3\gD(\vgL_7)=-261$;
  \item $\va^2=6$, i.e.\ $\va=\vr_i+\vr_j+\vr_k$
   for $i,j,k\neq0$ and $\vr_i\X\vr_j=\vr_j\X\vr_k=\vr_k\X\vr_i=0$, or
   $\va=2\vr_i+\vr_j$ for $i,j\neq0$ and $\vr_i\X\vr_j=-1$
   $\quad\Longrightarrow\quad
   [42\X(-1)^4+14\X(-1)^3]\gD(\vgL_0)=28$;
  \een
 \item $\vgL\X\va=-2$:
  \ben
  \item $\va^2=2$, i.e.\ $\va=\vr_0$
   $\quad\Longrightarrow\quad (-1)^2\gD(\vgL_0)=1$.
  \een
 \een
In total, this gives $-246+478-261+28+1=0$ for the simple
multiplicity. \\[1.5ex]
Let $\vgL=\vgL_7+\vgd$.
 \ben
 \item $\vgL\X\va=0$:
  \ben
  \item $\va=\vO$
   $\quad\Longrightarrow\quad -\gD(\vgL_7+\vgd)=-246$;
  \item $\va^2=2$, i.e.\ $\va=\vr_i$ for $i\neq-1,7$
   $\quad\Longrightarrow\quad
   8\X(-1)^2\gD(\vgL_1)=424$, \\
   since $\vgL-\va\in\fW(\vgL_1)$;
  \item $\va^2=4$, i.e.\ $\va=\vr_i+\vr_j$ for $i,j\neq -1,7$ and
   $\vr_i\X\vr_j=0$
   $\quad\Longrightarrow\quad
   21\X(-1)^3\gD(\vgL_7)=-189$;
  \item $\va^2=6$, i.e.\ $\va=\vr_i+\vr_j+\vr_k$ for $i,j,k\neq-1,7$
   and $\vr_i\X\vr_j=\vr_j\X\vr_k=\vr_k\X\vr_i=0$, or
   $\va=2\vr_i+\vr_j$ for $i,j\neq-1,7$ and $\vr_i\X\vr_j=-1$
   $\quad\Longrightarrow\quad
   [21\X(-1)^4+14\X(-1)^3]\gD(\vgL_0)=7$;
  \een
 \item $\vgL\X\va=-1$:
  \ben
  \item $\va^2=2$, i.e.\ $\va=\vrm$ or $\va=\vr_7$
   $\quad\Longrightarrow\quad
   2\X(-1)^2\gD(\vgL_7)=18$;
  \item $\va^2=4$, i.e.\ $\va=\vrm+\vr_i$ for $i\neq-1,0,7$ or
   $\va=\vr_7+\vr_j$ for $j\neq-1,6,7$ \\
   $\quad\Longrightarrow\quad
   14\X(-1)^3\gD(\vgL_0)=-14$.
  \een
 \een
In total, this gives $-246+424-189+7+18-14=0$ for the simple
multiplicity.
\item $\vgL^2=-10$, i.e.\ $\vgL=\vgL_1+\vgd$ or $\vgL=\vgL_8$. \\[1.5ex]
Let $\vgL=\vgL_1+\vgd$.
 \ben
 \item $\vgL\X\va=0$:
  \ben
  \item $\va=\vO$
   $\quad\Longrightarrow\quad
   -\gD(\vgL_1+\vgd)+(-1)^2\gD(\vgL_0)\gD(\vgL_0+\vr_0)=-982$, \\
   since $\vgL=2\vgL_0+\vr_0$;\footnote{Note that this is the first
   example where we have to take into account the product over the
   positive roots appearing on the l.h.s.\ of \Eq{Nenner3}.}
  \item $\va^2=2$, i.e.\ $\va=\vr_i$ for $i\neq-1,1$
   $\quad\Longrightarrow\quad
   8\X(-1)^2\gD(\vgL_7+\vgd)=1968$, \\
   since $\vgL-\vr_0\in\fW(2\vgL_0)$ and
   $\vgL-\vr_i\in\fW(\vgL_7+\vgd)$ for $i\neq-1,0,1$;
  \item $\va^2=4$, i.e.\ $\va=\vr_i+\vr_j$ for $i,j\neq -1,1$ and
   $\vr_i\X\vr_j=0$
   $\quad\Longrightarrow\quad
   22\X(-1)^3\gD(\vgL_1)=-1166$, \\
   since $\vgL-\va\in\fW(\vgL_1)$;
  \item $\va^2=6$, i.e.\ $\va=\vr_i+\vr_j+\vr_k$ for $i,j,k\neq-1,1$
   and $\vr_i\X\vr_j=\vr_j\X\vr_k=\vr_k\X\vr_i=0$, or
   $\va=2\vr_i+\vr_j$ for $i,j\neq-1,1$ and $\vr_i\X\vr_j=-1$
   $\quad\Longrightarrow\quad
   [26\X(-1)^4+12\X(-1)^3]\gD(\vgL_7)=126$;
  \item $\va^2=8$, i.e.\ $\va=\vr_i+\vr_j+\vr_k+\vr_l$
   for $i,j,k,l\neq-1,1$ and $\vr_i\X\vr_j=\ldots=\vr_l\X\vr_i=0$, or
   $\va=2\vr_i+\vr_j+\vr_k$ for $i,j,k\neq-1,1$ and $\vr_i\X\vr_j=-1$,
   $\vr_i\X\vr_k=\vr_j\X\vr_k=0$, or
   $\va=2(\vr_i+\vr_j)$ for $i,j\neq-1,1$ and $\vr_i\X\vr_j=-1$
   $\quad\Longrightarrow\quad
   [13\X(-1)^5+48\X(-1)^4+6\X(-1)^4]\gD(\vgL_0)=41$;
  \een
 \item $\vgL\X\va=-1$:
  \ben
  \item $\va^2=2$, i.e.\ $\va=\vrm$ or $\va=\vr_1$
   $\quad\Longrightarrow\quad
   (-1)^2[\gD(\vgL_0+2\vgd)+\gD(\vgL_1)]=107$, \\
   since $\vgL-\vrm\in\fW(\vgL_0+2\vgd)$
   and $\vgL-\vr_1\in\fW(\vgL_1)$;
  \item $\va^2=4$, i.e.\ $\va=\vrm+\vr_i$ for $i\neq-1,0,1$, or
   $\va=\vr_1+\vr_j$ for $j\neq-1,0,1,2$ \\
   $\quad\Longrightarrow\quad
   13\X(-1)^3\gD(\vgL_7)=-117$;
  \item $\va^2=6$, i.e.\ $\va=\vrm+2\vr_0$, $\va=\vr_1+2\vr_0$,
   $\va=\vr_1+2\vr_2$, or $\va=\vrm+\vr_i+\vr_j$ for $i,j\neq-1,0,1$
   and $\vr_i\X\vr_j=0$, or
   $\va=\vr_1+\vr_i+\vr_j$ for $i,j\neq-1,0,1,2$ and $\vr_i\X\vr_j=0$ \\
   $\quad\Longrightarrow\quad
   [25\X(-1)^4+3\X(-1)^3]\gD(\vgL_0)=22$;
  \een
 \item $\vgL\X\va=-2$:
  \ben
  \item $\va^2=4$, i.e.\ $\va=\vrm+\vr_1$
   $\quad\Longrightarrow\quad (-1)^3\gD(\vgL_0)=-1$.
  \een
 \een
In total, this gives $-982+1968-1166+126+41+107-117+22-1=-2$ contradicting
the conjecture of \ct{BaGeGuNi97}! We have to conclude that
$\vgL_1+\vgd$ is an imaginary simple root of multiplicity 2. \\[1.5ex]
Let $\vgL=\vgL_8$.
 \ben
 \item $\vgL\X\va=0$:
  \ben
  \item $\va=\vO$
   $\quad\Longrightarrow\quad
   -\gD(\vgL_8)=-981$;
  \item $\va^2=2$, i.e.\ $\va=\vr_i$ for $i\neq8$
   $\quad\Longrightarrow\quad
   9\X(-1)^2\gD(\vgL_7+\vgd)=2214$, \\
   since $\vgL-\va\in\fW(\vgL_7+\vgd)$;
  \item $\va^2=4$, i.e.\ $\va=\vr_i+\vr_j$ for $i,j\neq8$ and
   $\vr_i\X\vr_j=0$
   $\quad\Longrightarrow\quad
   28\X(-1)^3\gD(\vgL_1)=-1484$, \\
   since $\vgL-\va\in\fW(\vgL_1)$;
  \item $\va^2=6$, i.e.\ $\va=\vr_i+\vr_j+\vr_k$ for $i,j,k\neq8$
   and $\vr_i\X\vr_j=\vr_j\X\vr_k=\vr_k\X\vr_i=0$, or
   $\va=2\vr_i+\vr_j$ for $i,j\neq8$ and $\vr_i\X\vr_j=-1$
   $\quad\Longrightarrow\quad
   [35\X(-1)^4+16\X(-1)^3]\gD(\vgL_7)=171$;
  \item $\va^2=8$, i.e.\ $\va=\vr_i+\vr_j+\vr_k+\vr_l$
   for $i,j,k,l\neq8$ and $\vr_i\X\vr_j=\ldots=\vr_l\X\vr_i=0$, or
   $\va=2\vr_i+\vr_j+\vr_k$ for $i,j,k\neq8$ and $\vr_i\X\vr_j=-1$,
   $\vr_i\X\vr_k=\vr_j\X\vr_k=0$, or $\va=2(\vr_i+\vr_j)$ for
   $i,j\neq8$ and $\vr_i\X\vr_j=-1$
   $\quad\Longrightarrow\quad
   [15\X(-1)^5+84\X(-1)^4+8\X(-1)^4]\gD(\vgL_0)=77$;
  \een
 \item $\vgL\X\va=-1$:
  \ben
  \item $\va^2=2$, i.e.\ $\va=\vr_8$
   $\quad\Longrightarrow\quad
   (-1)^2\gD(\vgL_1)=53$, \\
   since $\vgL-\vr_8\in\fW(\vgL_1)$;
  \item $\va^2=4$, i.e.\ $\va=\vr_8+\vr_i$ for $i\neq5,8$
   $\quad\Longrightarrow\quad
   8\X(-1)^3\gD(\vgL_7)=-72$;
  \item $\va^2=6$, i.e.\ $\va=\vr_8+2\vr_5$ or $\va=\vr_8+\vr_i+\vr_j$
   for $i,j\neq5,8$ and $\vr_i\X\vr_j=0$ \\
   $\quad\Longrightarrow\quad
   [22\X(-1)^4+1\X(-1)^3]\gD(\vgL_0)=21$.
  \een
 \een
In total, this gives $-981+2214-1484+171+77+53-72+21=-1$ again
contradicting the conjecture. We conclude that $\vgL_8$ is an
imaginary simple root of multiplicity 1.
\een

%--------------------------------------------------------------------
\section{Simple Roots and Simple Multiplicities for $\vgL^2 \geq -24$}
%--------------------------------------------------------------------

The above calculations can now be carried much further with the help
of a computer, and in this section we present the results that we have
obtained down to norms $\vgL^2=-24$. Before giving these results, we
would, however, like to stress once more some of the extra complications
that arise as the root norms become more negative. As is already evident
from the example $\vgL=\vgL_1 + \vgd$ of the last section, we must
now deal with the product over all positive roots appearing on
the l.h.s.\ of \Eq{Nenner3}. More specifically, for a given root
$\vgL$ we will have to take into account all decompositions of
$\vv = \vgL - \va$ into sums of positive roots. To find them, we
make use of the following strategy. Since only positive roots with
nonvanishing $\gD$ contribute, we can disregard all real and
lightlike imaginary roots. Moreover, without loss of
generality we can rotate $\vv$ into the fundamental Weyl chamber
and look for decompositions $\vv = \sum_j \vv_j$ there; it is
important here that the summands $\vv_j$ need not belong to $\cC$
separately. Starting from any such decomposition, the action
of the little Weyl group $\cW(\vgL)$ (i.e., the stability subgroup
leaving $\vgL$ fixed) yields further decompositions. Hence we have to
take into account only those decompositions where at least one of the
elements $\vv_j$ is a lowest weight vector of $\cW(\vgL)$. In the cases we
have investigated we could restrict the search even further because only
decompositions into two components can occur. In general, however,
decompositions into an arbitrary number of positive roots will
have to be considered. For roots which are multiples of other
roots we also have another contribution, coming from higher terms
in the expansion of $(1-\re^\vv)^{\gD(\vv)}$.

Since in long computer calculations\footnote{For instance, the
determination of $\mu(\vgL_3)=2$ took four hours of CPU time.}
one can never exclude all possible sources of errors, we emphasize
once more that the positivity of the final result constitutes an
important consistency check, especially with descending norms as
the numbers involved in the sum become very large.
Our results are collected in Table 1.

\setlongtables
\doublerulesep 0.05ex
\begin{longtable}{|l|c|c|r|r|r|}
\caption{Simple multiplicities of imaginary simple roots\\
\protect\hphantom{Table 1: }with $\vgL^2 \le -2$}\\
\hline
\vphantom{\Large$M^M$}
$\vgL$&$\ell(\vgL)$ & ht$(\vgL)$ & $\vgL\!^2$ & $\dim\Lie9^{(\vgL)}$ & $\gm(\vgL)$  \\
\hline
\hline \endhead
\hline \endfoot
$\vgL_0              $ & 1 & 61  & -2  &         45 &  1\\
$\vgL_7              $ & 2 & 76  & -4  &        201 &  0\\
$\vgd+\vgL_0         $ & 1 & 91  & -4  &        201 &  1\\
$\vgL_1              $ & 2 & 93  & -6  &        780 &  0\\
$\vgd+\vgL_7         $ & 2 & 106 & -8  &       2718 &  0\\
$\vgL_8              $ & 3 & 115 & -10 &       8730 &  1\\
$2\vgd+\vgL_0        $ & 1 & 121 & -6  &        780 &  2\\
$2\vgL_0             $ & 2 & 122 & -8  &       2718 &  0\\
$\vgd+\vgL_1         $ & 2 & 123 & -10 &       8730 &  2\\
$\vgL_{2}            $ & 3 & 126 & -12 &      26226 &  0\\
$2\vgd+\vgL_7        $ & 2 & 136 & -12 &      26226 &  1\\
$\vgL_0+\vgL_7       $ & 3 & 137 & -14 &      74556 &  3\\
$\vgd+\vgL_8         $ & 3 & 145 & -16 &     202180 &  3\\
$3\vgd+\vgL_0        $ & 1 & 151 & -8  &       2718 &  2\\
$\vgd+2\vgL_0        $ & 2 & 152 & -12 &      26226 &  2\\
$2\vgL_7             $ & 4 & 152 & -16 &     202180 &  0\\
$2\vgd+\vgL_1        $ & 2 & 153 & -14 &      74556 &  3\\
$\vgL_{6}            $ & 4 & 153 & -18 &     526397 &  3\\
$\vgL_0+\vgL_1       $ & 3 & 154 & -16 &     202180 &  4\\
$\vgd+\vgL_{2}       $ & 3 & 156 & -18 &     526397 &  6\\
$\vgL_{3}            $ & 4 & 160 & -20 &    1322343 &  2\\
$3\vgd+\vgL_7        $ & 2 & 166 & -16 &     202180 &  4\\
$\vgd+\vgL_0+\vgL_7  $ & 3 & 167 & -20 &    1322343 & 14\\
$\vgL_1+\vgL_7       $ & 4 & 169 & -22 &    3218091 & 14\\
$2\vgd+\vgL_8        $ & 3 & 175 & -22 &    3218091 & 14\\
$\vgL_0+\vgL_8       $ & 4 & 176 & -24 &    7612014 &   \\
$4\vgd+\vgL_0        $ & 1 & 181 & -10 &       8730 &  4\\
$2\vgd+2\vgL_0       $ & 2 & 182 & -16 &     202180 &  4\\
$\vgd+2\vgL_7        $ & 4 & 182 & -24 &    7612014 &   \\
$3\vgd+\vgL_1        $ & 2 & 183 & -18 &     526397 & 11\\
$3\vgL_0             $ & 3 & 183 & -18 &     526397 &  7\\
$\vgd+\vgL_{6}       $ & 4 & 183 & -26 &   17548920 &   \\
%%%%%%%%%%%%%%%%%%%%%%%%%%%%%%%%%%%%%%%%%%%%%%%%%%%%%%%%%%
% Borcherds-Multiplizitaeten bis hierher richtig eingetragen!
% und bei den auskommentierten unteren Zeilen!
%%%%%%%%%%%%%%%%%%%%%%%%%%%%%%%%%%%%%%%%%%%%%%%%%%%%%%%%%%
%$\vgd+\vgL_0+\vgL_1  $ & 3 & 184 & -22 &    2483871 &   \\
%$2\vgd+\vgL_{2}      $ & 3 & 186 & -24 &    5746226 &   \\
%$2\vgL_1             $ & 4 & 186 & -24 &    5749565 &   \\
%$\vgL_0+\vgL_{2}     $ & 4 & 187 & -26 &   12970045 &   \\
%$\vgd+\vgL_{3}       $ & 4 & 190 & -28 &   28592513 &   \\
%$\vgL_7+\vgL_8       $ & 5 & 191 & -28 &   28595548 &   \\
%$\vgL_{4}            $ & 5 & 195 & -30 &   61721165 &   \\
$4\vgd+\vgL_7         $ & 2 & 196 & -20 &    1322343 & 8 \\
%$2\vgd+\vgL_0+\vgL_7 $ & 3 & 197 & -26 &   12959290 &   \\
%$2\vgL_0+\vgL_7      $ & 4 & 198 & -28 &   28589025 &   \\
%$\vgd+\vgL_1+\vgL_7  $ & 4 & 199 & -30 &   61711591 &   \\
%$\vgL_{2}+\vgL_7     $ & 5 & 202 & -32 &  130661924 &   \\
%$3\vgd+\vgL_8        $ & 3 & 205 & -28 &   28559052 &   \\
%$\vgd+\vgL_0+\vgL_8  $ & 4 & 206 & -32 &  130632964 &   \\
%$\vgL_1+\vgL_8       $ & 5 & 208 & -34 &  271695444 &   \\
%$5\vgd+\vgL_0        $ & 1 & 211 & -12 &      22528 &   \\
$3\vgd+2\vgL_0        $ & 2 & 212 & -20 &    1322343 & 13\\
%$2\vgd+2\vgL_7       $ & 4 & 212 & -32 &  130635596 &   \\
$4\vgd+\vgL_1         $ & 2 & 213 & -22 &    3218091 & 25\\
%$\vgd+3\vgL_0        $ & 3 & 213 & -24 &    5745720 &   \\
%$\vgL_0+2\vgL_7      $ & 5 & 213 & -34 &  271702532 &   \\
%$2\vgd+\vgL_{6}      $ & 4 & 213 & -34 &  271618575 &   \\
%$2\vgd+\vgL_0+\vgL_1 $ & 3 & 214 & -28 &   28558597 &   \\
%$\vgL_0+\vgL_6       $ & 5 & 214 & -36 &  555652661 &   \\
%$2\vgL_0+\vgL_1      $ & 4 & 215 & -30 &   61699285 &   \\
%$3\vgd+\vgL_{2}      $ & 3 & 216 & -30 &   61620301 &   \\
%$\vgd+2\vgL_1        $ & 4 & 216 & -32 &  130630342 &   \\
%$\vgd+\vgL_0+\vgL_2  $ & 4 & 217 & -34 &  271609694 &   \\
%$\vgL_1+\vgL_{2}     $ & 5 & 219 & -36 &  555631102 &   \\
%$2\vgd+\vgL_{3}      $ & 4 & 220 & -36 &  555434128 &   \\
%$\vgd+\vgL_7+\vgL_8  $ & 5 & 221 & -38 & 1118955631 &   \\
%$\vgL_0+\vgL_{3}     $ & 5 & 221 & -38 & 1118894437 &   \\
%$\vgd+\vgL_{4}       $ & 5 & 225 & -40 & 2220872914 &   \\
%$5\vgd+\vgL_7        $ & 2 & 226 & -24 &    5717880 &   \\
%$3\vgd+\vgL_0+\vgL_7 $ & 3 & 227 & -32 &  130395100 &   \\
%$\vgd+2\vgL_0+\vgL_7 $ & 4 & 228 & -36 &  555404364 &   \\
%$3\vgL_7             $ & 6 & 228 & -36 &  555695680 &   \\
%$2\vgd+\vgL_1+\vgL_7 $ & 4 & 229 & -38 & 1118347860 &   \\
%$\vgL_{6}+\vgL_7     $ & 6 & 229 & -40 & 2221039540 &   \\
%$\vgL_0+\vgL_1+\vgL_7$ & 5 & 230 & -40 & 2220699951 &   \\
%$2\vgL_8             $ & 6 & 230 & -40 & 2221026189 &   \\
%$\vgL_5              $ & 6 & 231 & -42 & 4348985101 &   \\
\hline
\end{longtable}

For the convenience of the reader and to provide a ``bird's eye's ''
view on the results obtained so far, we have displayed them
once more in the table below. This table highlights two facts, namely,
(i) that in some cases the simple multiplicities depend only on
the norm of the root in question, and (ii) (somewhat to our surprise)
that the simple multiplicities at level $\cl =2$ depend also on the
``direction'' of the root, unlike the corresponding \0 multiplicities!

\renewcommand{\thefootnote}{\fnsymbol{footnote}}
\setcounter{footnote}{0}
\begin{table}[htb]
\begin{center}
\begin{tabular}{c|cccccccccccc|c}
$\cl\big\backslash \vgL^2$\!\!\!\!
     &  -2 &  -4 &  -6 &  -8 & -10 & -12 & -14 & -16 & -18 & -20 & -22 & -24\\ \hline
  1  &   1 &   1 &   2 &   2 &   4 &   4 &   7 &   8 &  12 &  14 &  21 &  24\\
  2  &     &   0 &   0 & 0\footnotemark &   2 & 1,2 &   3 & $4^\thefootnote$ &  11 & 8,13&  25 &    \\
  3  &     &     &     &     &   1 &  0  &   3 & 3,4 & 6,7 &  14 &  14 &    \\
  4  &     &     &     &     &     &     &     &   0 &   3 &   2 &  14 &    \\
\hline
\end{tabular}
\caption{Simple multiplicities of imaginary simple roots for \Lie9\lb{fig-1}}
\end{center}
\end{table}
\footnotetext{Occurs twice.}
While no clear pattern is discernible in the simple root multiplicities
so far, the smallness of the numbers obtained is noteworthy.
Especially the zeroes in this table (the corresponding roots thus must
not be counted as simple roots) appear to us quite striking in view of the
fact that we are unaware of any obvious {\it a priori} reason for
their existence. Furthermore, we observe that the simple multiplicities
do not depend monotonically on the norms for levels $\cl \geq 2$,
unlike the level-1 simple multiplicities, and unlike the simple
multiplicities of the gnome Lie algebra. Also, for fixed $\vgL^2$,
there is a tendency for the simple multiplicities to decrease
as a function of the level. Of course, these results might just
be a coincidence, and the simple multiplicities, although rather well
behaved for low norms and low levels, might explode after a
few more steps. On the other hand, the fact that we stay so close
to zero makes us wonder if there is not a hidden structure
``just around the corner.''

The smallness of the simple root multiplicities means that
that \0 is a rather ``big'' subalgebra of $\Lie9$. This behavior
is to be contrasted with that of the gnome Lie algebra $\Lie1$,
whose maximal Kac Moody subalgebra is the finite algebra $sl(2,\Rn)$.
However, readers should keep in mind that \Eq{E10} admits
an infinite nested sequence of Borcherds algebras ``between''
\0 and $\Lie9$: these are simply obtained obtained by omitting
any number of missing \0 modules from $\Lie9$ or, equivalently,
the corresponding simple roots from the root system of $\Lie9$.

Being confronted with the ineluctable conclusion that \0 is much
more complicated than either the gnome or the fake monster, the next
question is, how should one proceed from here onwards?
Recall that the nice structure underlying the root system of the
fake monster Lie algebra was discovered by methods very similar to the
ones employed here (see remarks in Sect.\ 5 of \ct{Borc88}). By
computing the simple multiplicities of roots down to norm $-6$,
Borcherds realized that the imaginary simple roots are all proportional
to $\vgr$ with uniform multiplicity 24 (corresponding to the 24
transverse polarizations of a photon in 26 dimensions), where $\vgr$
is the lightlike Weyl vector of $\latt{25}$; and, happily,
the pattern thus reveals itself after only very few steps!
(Observe that, in fact, for the fake monster, all entries in Table 2
would vanish because $\Lie{25}$ has no simple roots of negative norm.)
For $\0$, we are evidently not in such a fortuitous situation,
and at this point the only feasible way to make further progress with
presently available techniques seems to be to collect even more data
about the simple multiplicities.  Fortunately, we have seen that our
method can be conveniently implemented on a computer.

Assuming a general pattern for the simple multiplicities
we would still face the problem of a rigorous proof.
For the monster, Borcherds was able to prove that the emerging
pattern was, in fact, a general property of the algebra and its root
system by establishing a new modular identity. In the case at hand,
the question is therefore whether our new denominator formula
\Eq{Nenner3} admits a modular interpretation, too. This question is
obviously of a more general interest, as similar modified denominator
formulas are expected to exist for other algebras of this type.
After a suitable specialization, these formulas would give
rise to new modular identities involving all levels
simultaneously. In making these speculations,
we are encouraged by the fact that there do exist examples
of automorphic forms which give rise to Borcherds algebras
with \0 as maximal Kac--Moody subalgebra (see Example 1 in
Sect.\ 16 of \ct{Borc95a}, and \ct{HarMoo96}).

\medskip

\noindent{\bf Acknowledgments:} We are indebted to
R.~Borcherds for sharing with us his (unpublished) expertise on
reflection groups and the Peterson formula, and to J.~Fuchs
for explaining to us his Weyl orbit method. H.~N. would like to
thank the Newton Institute in Cambridge, where part of this work
was carried out, for hospitality and support. O.~B. is grateful
to the Albert-Einstein-Institut for hospitality during a visit there.

\begin{appendix}
\section{$E_{10}$ Multiplicities for $\vgL\in\cC$ with ht($\vgL) \le 231$}
In this appendix we collect the multiplicities of all roots of \0
with height $\le 231$. This includes the multiplicities of all
fundamental weights of \0.

The calculation of these multiplicities starts from Peterson's
formula, which in principle allows the recursive computation of the
multiplicity of any root. For algebras of high rank such as \0,
however, this procedure soon takes up too much time due to the large
number of roots involved. We will use an approach due to Borcherds
\ct{Borcherds} to simplify the calculations.  The idea is to employ
the little Weyl (or stability) group of the root in question
and to group the roots into orbits of this group and then count these
orbits rather than the roots themselves. One has the following identity

\[
(\vgL | \vgL - 2\vgr)c_\vgL=
\sum_{
\begin{minipage}{2cm}
\begin{center}
\scriptsize$\vgb', \vgb''\in Q_+$\\$\vgL = \vgb' + \vgb''$
\end{center}
\end{minipage}
}(\vgb'| \vgb '') c_{\vgb'}c_{\vgb''}
\stackrel{!}{=} \sum_{\vv \in Q_+}(\vv|\vgL-\vv)c_{\vv}c_{\vgL-\vv}|\cW(\vv)|.
\]
The second sum is over all real roots and all lowest weight vectors
$\vv$ in $Q_+$ with respect to the little Weyl group $\cW(\vgL)$ such
that $\vgL-\vv$ is also a positive root. One also has to be careful not to count the
same orbit twice. $|\cW(\vv)|$ denotes the size of the orbit of
this lowest weight vector under $\cW(\vgL)$ and we have
\[
c_\vgb=\sum_{k \ge 1} \frac{1}{k}\mult\left(\frac{\vgb}{k}\right).
\]

To give an example for this procedure we consider the two simplest cases:
$\vgL=\vgL_{-1}=\vgd$ and $\vgL=\vgL_{0}$.
\begin{enumerate}
\item $\vgL=\vgd \Rightarrow (\vgL | \vgL - 2\vgr)=-60$
\begin{enumerate}
\item $\vgd=\vr_0 + \vgt \Rightarrow (\vgt|\vgL-\vgt)c_{\vgt}c_{\vgL-\vgt}|\cW(\vgt)|=-240$
\end{enumerate}
Taking an extra factor of 2 due to the symmetry of the sum
into account we recover the well-known result $\mult(\vgd)=8$.
\item $\vgL=\vgL_0 \Rightarrow (\vgL | \vgL - 2\vgr)=-124$
\begin{enumerate}
\item $\vgL_0=\vgd     + (\vgL_0-\vgd)
\Rightarrow (\vgd|\vgL-\vgd)c_{\vgd}c_{\vgL-\vgd}|\cW(\vgd)|=-64$
\item $\vgL_0=\vr_{-1} + (\vgL_0-\vr_{-1})
\Rightarrow(\vr_{-1}|\vgL-\vr_{-1})c_{\vr_{-1}}c_{\vgL-\vr_{-1}}|\cW(\vr_{-1})|= -24$
\item $\vgL_0=\vr_0    + (\vgL_0-\vr_0)
\Rightarrow(\vr_0|\vgL-\vr_0)c_{\vr_0}c_{\vgL-\vr_0}|\cW(\vr_0)|= -1920$
\item $\vgL_0=\vr_1    + (\vgL_0-\vr_1)
\Rightarrow(\vr_1|\vgL-\vr_1)c_{\vr_1}c_{\vgL-\vr_1}|\cW(\vr_1)|= -720$
\end{enumerate}
Taking an extra factor of 2 due to the symmetry of the sum
into account we find the expected result $\mult(\vgL_0)=44$.
\end{enumerate}

For a given $\vv$ the size of the Weyl orbit is easily calculated as
follows: For $\vgL \neq \vgd$ the little Weyl group is a finite group
with known order.  We can also assume that $\vv$ is a lowest weight
vector for this group. The subgroup fixing it is then the Weyl group
whose simple roots are those orthogonal to the vector $\vv$. The order
of this subgroup is calculated by looking it up, determining the
simple factors from the Dynkin diagram. The size of the orbit is then
given as the quotient of these two orders.

The remaining problem is to find all lowest weight
vectors for $\cW(\vgL)$ in $Q_+$. These are given by all positive roots of the form
\[
\vv = \sum_{i=-1}^8 n_i \vgL_i
\]
where only the coefficients $n_i$ corresponding to simple roots
$\vr_i$ orthogonal to $\vgL$ have to be positive. Due to the symmetry
of the sum one can restrict the search to vectors with heights ht$(\vv) \le [$ht$(\vgL)/2]$.

For roots of low height this formula can be evaluated by hand but this
becomes impractical very quickly due to the large number of orbits.
For $\vgL=\vgL_0 +\vgL_1 + \vgL_7$ e.g.\ there are 635 contributing
Weyl orbits.  Hence most of the multiplicities were calculated with a
computer using the symbolic algebra system {\tt Maple V}. One
consistency check here is that, despite the occurrence of fractional
numbers at intermediate stages of the calculation, the final result
must be an integer.

Our results for the \0 multiplicities and the values of
$\gD$ (cf.\ \Eq{D}) are collected in Table 3 below, where we have
labeled the root $\vgL = \sum_{j=-1}^8 n_j \vgL_j$ by the
symbol $[n_{-1},n_0,n_1,\dots ,n_8]$.

%\setlongtables
\doublerulesep 0.05ex
\begin{longtable}{|l|l|c|r|r|r|r|}
\caption{$E_{10}$ Root Multiplicities}\\
\hline
\vphantom{\Large $M^M$}$\vgL$& $\vgL$ & $\ell(\vgL)$ & ht$(\vgL)$ & $\vgL\!^2$      &mult$(\vgL)$ & $\gD(\vgL)$\\
\hline
\hline \endhead
\hline \endfoot
$\vgd              $ & $[0, 1, 2, 3, 4, 5, 6, 4, 2, 3]$ & 0 & 30 & 0 & 8 & 0 \\
$2\vgd             $ & $[0, 2, 4, 6, 8, 10, 12, 8, 4, 6]$ & 0 & 60 & 0 &  8 & 0\\
$\vgL_{0}          $ & $[1, 2, 4, 6, 8, 10, 12, 8, 4, 6]$ & 1 & 61 & -2 & 44  & 1\\
$\vgL_{7}          $ & $[2, 4, 6, 8, 10, 12, 14, 9, 4, 7]$ & 2 & 76 & -4 & 19 2 & 9\\
$3\vgd             $ & $[0, 3, 6, 9, 12, 15, 18, 12, 6, 9]$ & 0 & 90 & 0  & 8 & 0\\
$\vgd+\vgL_{0}     $ & $[1, 3, 6, 9, 12, 15, 18, 12, 6, 9]$ & 1  & 91 & -4 & 192 & 9\\
$\vgL_{1}          $ & $[2, 4, 6, 9, 12, 15, 18, 12, 6, 9]$ & 2 & 93 & -6 & 7 27 & 53\\
$\vgd+\vgL_7       $ & $[2, 5, 8, 11, 14, 17, 20, 13, 6, 10]$ &  2 & 106 & -8 & 2472 & 246\\
$\vgL_{8}          $ & $[3, 6, 9, 12, 15, 18, 21, 14, 7, 10]$ & 3 & 115 & -10  & 7749 & 981\\
$4\vgd             $ & $[0, 4, 8, 12, 16, 20, 24, 16, 8, 12]$ & 0 & 120  & 0 & 8 & 0\\
$2\vgd+\vgL_0      $ & $[1, 4, 8, 12, 16, 20, 24, 16, 8, 12]$  & 1 & 121 & -6 & 726 & 54\\
$2\vgL_{0}         $ & $[2, 4, 8, 12, 16, 20, 24, 16, 8, 12]$ & 2 & 122 &  -8 & 2472 & 246\\
$\vgd+\vgL_{1}     $ & $[2, 5, 8, 12, 16, 20, 24, 16, 8, 12]$ &  2 & 123 & -10 & 7747 & 983\\
$\vgL_{2}          $ & $[3, 6, 9, 12, 16, 20, 24, 16, 8, 12]$ & 3 & 126 & -12  & 22725 & 3501\\
$2\vgd+\vgL_7      $ & $[2, 6, 10, 14, 18, 22, 26, 17, 8, 13]$  & 2 & 136 & -12 & 22712 & 3514\\
$\vgL_{0}+\vgL_{7} $ & $[3, 6, 10, 14, 18, 22, 26, 17, 8, 13]$ &  3 & 137 & -14 & 63085 & 11471\\
$\vgd+\vgL_{8}     $ & $[3, 7, 11, 15, 19, 23, 27, 18, 9, 13]$ &  3 & 145 & -16 & 167116 & 35064\\
$5\vgd             $ & $[0, 5, 10, 15, 20, 25, 30, 20, 10, 15]$ & 0 & 150 & 0 & 8 & 0\\
$3\vgd+\vgL_0      $ & $[1, 5, 10, 15, 20, 25, 30, 20, 10, 15] $ & 1 & 151 & -8 & 2464 & 254\\
$\vgd+2\vgL_0      $ & $[2, 5, 10, 15, 20, 25, 30, 20, 10, 15] $ & 2 & 152 & -12 & 22712 & 3514\\
$2\vgL_{7}         $ & $[4, 8, 12, 16, 20, 24, 28, 18, 8, 14]$ & 4 & 152  & -16 & 167133 & 35047\\
$2\vgd+\vgL_1      $ & $[2, 6, 10, 15, 20, 25, 30, 20, 10, 15] $ & 2 & 153 & -14 & 63020 & 11536\\
$\vgL_{6}          $ & $[4, 8, 12, 16, 20, 24, 28, 18, 9, 14]$ & 4 & 153 & -1 8 & 425227 & 101170\\
$\vgL_{0}+\vgL_{1} $ & $[3, 6, 10, 15, 20, 25, 30, 20, 10, 15]$ &  3 & 154 & -16 & 167099 & 35081\\
$\vgd+\vgL_{2}     $ & $[3, 7, 11, 15, 20, 25, 30, 20, 10, 15]$  & 3 & 156 & -18 & 425156 & 101241\\
$\vgL_{3}          $ & $[4, 8, 12, 16, 20, 25, 30, 20, 10, 15]$ & 4 & 160 & - 20 & 1044218 & 278125\\
$3\vgd+\vgL_7      $ & $[2, 7, 12, 17, 22, 27, 32, 21, 10, 16] $ & 2 & 166 & -16 & 166840 & 35340\\
$\vgd+\vgL_0+\vgL_7$ & $[3, 7, 12, 17, 22, 27, 32, 21,  10, 16]$ & 3 & 167 & -20 & 1043926 & 278417\\
$\vgL_{1}+\vgL_{7} $ & $[4, 8, 12, 17, 22, 27, 32, 21, 10, 16]$ &  4 & 169 & -22 & 2485020 & 733071\\
$2\vgd+\vgL_8      $ & $[3, 8, 13, 18, 23, 28, 33, 22, 11, 16] $ & 3 & 175 & -22 & 2483970 & 734121\\
$\vgL_{0}+\vgL_{8} $ & $[4, 8, 13, 18, 23, 28, 33, 22, 11, 16]$ &  4 & 176 & -24 & 5749818 & 1862196\\
$6\vgd             $ & $[0, 6, 12, 18, 24, 30, 36, 24, 12, 18]$ & 0 & 180 & 0 & 8 & 0\\
$4\vgd+\vgL_{0}    $ & $[1, 6, 12, 18, 24, 30, 36, 24, 12, 18] $ & 1 & 181 & -10 & 7704 & 1026\\
$2\vgd+2\vgL_{0}   $ & $[2, 6, 12, 18, 24, 30, 36, 24, 12, 18]$ & 2 & 182 & -16 & 166840 & 35340\\
$\vgd+2\vgL_{7}    $ & $[4, 9, 14, 19, 24, 29, 34, 22, 10, 17] $ & 4 & 182 & -24 & 5750072 & 1861942\\
$3\vgd+\vgL_{1}    $ & $[2, 7, 12, 18, 24, 30, 36, 24, 12, 18] $ & 2 & 183 & -18 & 424161 & 102236\\
$3\vgL_{0}         $ & $[3, 6, 12, 18, 24, 30, 36, 24, 12, 18]$ & 3 & 183  & -18 & 425058 & 101339\\
$\vgd+\vgL_{6}     $ & $[4, 9, 14, 19, 24, 29, 34, 22, 11, 17]$  & 4 & 183 & -26 & 12971009 & 4577911\\
$\vgd+\vgL_0+\vgL_1$ & $[3, 7, 12, 18, 24, 30, 36, 24,  12, 18]$ & 3 & 184 & -22 & 2483871 & 734220\\
$2\vgd+\vgL_{2}    $ & $[3, 8, 13, 18, 24, 30, 36, 24, 12, 18] $ & 3 & 186 & -24 & 5746226 & 1865788\\
$2\vgL_{1}         $ & $[4, 8, 12, 18, 24, 30, 36, 24, 12, 18]$ & 4 & 186  & -24 & 5749565 & 1862449\\
$\vgL_{0}+\vgL_{2} $ & $[4, 8, 13, 18, 24, 30, 36, 24, 12, 18]$ &  4 & 187 & -26 & 12970045 & 4578875\\
$\vgd+\vgL_{3}     $ & $[4, 9, 14, 19, 24, 30, 36, 24, 12, 18]$  & 4 & 190 & -28 & 28592513 & 10931086\\
$\vgL_{7}+\vgL_{8} $ & $[5, 10, 15, 20, 25, 30, 35, 23, 11, 17]$  & 5 & 191 & -28 & 28595548 & 10928051\\
$\vgL_{4}          $ & $[5, 10, 15, 20, 25, 30, 36, 24, 12, 18]$ & 5 & 195 &  -30 & 61721165 & 25411831\\
$4\vgd+\vgL_{7}    $ & $[2, 8, 14, 20, 26, 32, 38, 25, 12, 19] $ & 2 & 196 & -20 & 1040664 & 281679\\
$2\vgd+\vgL_0+\vgL_7$ & $[3, 8, 14, 20, 26, 32, 38, 2 5, 12, 19]$ & 3 & 197 & -26 & 12959290 & 4589630\\
$2\vgL_{0}+\vgL_{7}$ & $[4, 8, 14, 20, 26, 32, 38, 25, 12, 19]$  & 4 & 198 & -28 & 28589025 & 10934574\\
$\vgd+\vgL_1+\vgL_7$ & $[4, 9, 14, 20, 26, 32, 38, 25,  12, 19]$ & 4 & 199 & -30 & 61711591 & 25421405\\
$\vgL_{2}+\vgL_{7} $ & $[5, 10, 15, 20, 26, 32, 38, 25, 12, 19]$  & 5 & 202 & -32 & 130661924 & 57690454\\
$3\vgd+\vgL_{8}    $ & $[3, 9, 15, 21, 27, 33, 39, 26, 13, 19] $ & 3 & 205 & -28 & 28559052 & 10964547\\
$\vgd+\vgL_{0}+\vgL_{8}$ & $[4, 9, 15, 21, 27, 33, 39, 26,  13, 19]$ & 4 & 206 & -32 & 130632964 & 57719414\\
$\vgL_{1}+\vgL_{8}$ & $[5, 10, 15, 21, 27, 33, 39, 26, 13, 19]$  & 5 & 208 & -34 & 271695444 & 128129588\\
$7\vgd$ & $[0, 7, 14, 21, 28, 35, 42, 28, 14, 21]$ & 0 & 210 & 0 & 8 & 0\\
$5\vgd+\vgL_{0}$ & $[1, 7, 14, 21, 28, 35, 42, 28, 14, 21] $ & 1 & 211 & -12 & 22528 & 3698\\
$3\vgd+2\vgL_{0}$ & $[2, 7, 14, 21, 28, 35, 42, 28, 14, 21]$ & 2 & 212 & -20 & 1040664 & 281679\\
$2\vgd+2\vgL_{7}$ & $[4, 10, 16, 22, 28, 34, 40, 26, 12,  20]$ & 4 & 212 & -32 & 130635596 & 57716782\\
$4\vgd+\vgL_{1}$ & $[2, 8, 14, 21, 28, 35, 42, 28, 14, 21] $ & 2 & 213 & -22 & 2474026 & 744065\\
$\vgd+3\vgL_{0}$ & $[3, 7, 14, 21, 28, 35, 42, 28, 14, 21] $ & 3 & 213 & -24 & 5745720 & 1866294\\
$\vgL_{0}+2\vgL_{7}$ & $[5, 10, 16, 22, 28, 34, 40, 26, 12, 20] $ & 5 & 213 & -34 & 271702532 & 128122500\\
$2\vgd+\vgL_{6}$ & $[4, 10, 16, 22, 28, 34, 40, 26, 13, 20]$ & 4 & 213 & -34 & 271618575 & 128206457\\
$2\vgd+\vgL_{0}+\vgL_{1}$ & $[3, 8, 14, 21, 28, 35, 42, 2 8, 14, 21]$ & 3 & 214 & -28 & 28558597 & 10965002\\
$\vgL_{0}+\vgL_{6}$ & $[5, 10, 16, 22, 28, 34, 40, 26, 13, 20]$ & 5 & 214 & -36 & 555652661 & 278885204\\
$2\vgL_{0}+\vgL_{1}$ & $[4, 8, 14, 21, 28, 35, 42, 28, 14, 21]$  & 4 & 215 & -30 & 61699285 & 25433711\\
$3\vgd+\vgL_{2}$ & $[3, 9, 15, 21, 28, 35, 42, 28, 14, 21] $ & 3 & 216 & -30 & 61620301 & 25512695\\
$\vgd+2\vgL_{1}$ & $[4, 9, 14, 21, 28, 35, 42, 28, 14, 21] $ & 4 & 216 & -32 & 130630342 & 57722036\\
$\vgd+\vgL_{0}+\vgL_{2}$ & $[4, 9, 15, 21, 28, 35, 42, 28,  14, 21]$ & 4 & 217 & -34 & 271609694 & 128215338\\
$\vgL_{1}+\vgL_{2}$ & $[5, 10, 15, 21, 28, 35, 42, 28, 14, 21]$  & 5 & 219 & -36 & 555631102 & 278906763\\
$2\vgd+\vgL_{3}$ & $[4, 10, 16, 22, 28, 35, 42, 28, 14, 21]$ & 4 & 220 & -36 & 555434128 & 279103737\\
$\vgd+\vgL_{7}+\vgL_{8}$ & $[5, 11, 17, 23, 29, 35, 41, 27 , 13, 20]$ & 5 & 221 & -38 & 1118955631 & 595793867\\
$\vgL_{0}+\vgL_{3}$ & $[5, 10, 16, 22, 28, 35, 42, 28, 14, 21]$  & 5 & 221 & -38 & 1118894437 & 595855061\\
$\vgd+\vgL_{4}$ & $[5, 11, 17, 23, 29, 35, 42, 28, 14, 21]$  & 5 & 225 & -40 & 2220872914 & 1251118823\\
$5\vgd+\vgL_{7}$ & $[2, 9, 16, 23, 30, 37, 44, 29, 14, 22] $ & 2 & 226 & -24 & 5717880 & 1894134\\
$3\vgd+\vgL_{0}+\vgL_{7}$ & $[3, 9, 16, 23, 30, 37, 44, 2 9, 14, 22]$ & 3 & 227 & -32 & 130395100 & 57957278\\
$\vgd+2\vgL_{0}+\vgL_{7}$ & $[4, 9, 16, 23, 30, 37, 44, 2 9, 14, 22]$ & 4 & 228 & -36 & 555404364 & 279133501\\
$3\vgL_{7}$ & $[6, 12, 18, 24, 30, 36, 42, 27, 12, 21]$ & 6 & 228 & -36 & 555695680 & 278842185\\
$2\vgd+\vgL_{1}+\vgL_{7}$ & $[4, 10, 16, 23, 30, 37, 44,  29, 14, 22]$ & 4 & 229 & -38 & 1118347860 & 596401638\\
$\vgL_{6}+\vgL_{7}$ & $[6, 12, 18, 24, 30, 36, 42, 27, 13, 21]$  & 6 & 229 & -40 & 2221039540 & 1250952197\\
$\vgL_{0}+\vgL_{1}+\vgL_{7}$ & $[5, 10, 16, 23, 30, 37, 44, 29,  14, 22]$ & 5 & 230 & -40 & 2220699951 & 1251291786\\
$2\vgL_{8}$ & $[6, 12, 18, 24, 30, 36, 42, 28, 14, 20]$ & 6 & 230 & -40 & 2221026189 & 1250965548\\
$\vgL_{5}$ & $[6, 12, 18, 24, 30, 36, 42, 28, 14, 21]$ & 6 & 231 &  -42 & 4348985101 & 2584919075\\
\hline
\end{longtable}
\end{appendix}

%\bibliography{Liste}
%\bibliographystyle{Springer}

\end{document}